\documentclass[twocolumn,showpacs,groupedaddress]{revtex4}
\usepackage{graphicx}
\usepackage{amsmath}

\begin{document}

\title{Hall conductance and topological invariant for open systems}
\author{H. Z. Shen$^{1,2}$, W. Wang$^1$ and   X. X. Yi$^1$ }

\affiliation{$^1$ Center for Quantum Sciences and School of Physics,
Northeast Normal University, Changchun 130024, China\\
$^2$ School of Physics and Optoelectronic Technology, Dalian
University of Technology, Dalian 116024, China }

\date{\today}

\begin{abstract}
The Hall conductivity given by the Kubo formula is a linear response
of the quantum transverse transport to a weak electric field. It has
been intensively studied for a quantum  system without decoherence,
but it is barely explored for systems subject to decoherence. In
this paper, we develop a formulism to deal with this issue for
topological insulators. The Hall conductance for a topological
insulator coupled to an environment is derived, the derivation is
based on a linear response theory of open system. As an application,
the Hall conductance of a two-band topological insulator and a
two-dimensional lattice  is presented and discussed.
\end{abstract}

 \maketitle

Topological insulators (TIs) were theoretically predicted to exist
and have been experimentally discovered in
\cite{konig07,hsieh08,roth09}, they are materials that have a bulk
electronic band gap like an ordinary insulator but have protected
conducting topological states(edge states)  on their surface. In the
last decades, these topological materials  have gained many
interests of scientific community for their unique properties such
as quantized conductivities, dissipationless transport and edge
states physics\cite{hasan10,qi11}. Although the exploration of
topological phases of matter has become a major topics at the
frontiers of the condensed matter physics, the behavior of TIs
subject to dissipative dynamics has been barely explored. This leads
to  a lack of capability to discuss issues such as their robustness
against decoherence, which is crucial in applications of the
materials in quantum information processing and spintronics.

Most recently, the study of  topological states was extended to
non-unitary systems\cite{viyuela14,rivas13,viyuela86155140}, going a
step further beyond the Hamiltonian ground-state scenario. This
first step  was taken with specifically designed dissipative
dynamics described by a quantum master equation. Such an approach
was originally proposed as a means of quantum state preparation and
quantum computation\cite{yi03}, which  relies on the engineering of
the system-reservoir coupling. To define the topological invariant
for open systems, the authors  use  a scheme called purification  to
calculate quantities of quantum system in mixed states. To be
specific, for a density matrix $\rho$  in a Hilbert space
$\mathcal{H}$, the density matrix $\rho$ can be purified to
 $|{\Phi^\rho}\rangle$ by introducing an ancilla
acting on a Hilbert space $\mathcal{H}_A$ such that the tracing over
the ancilla $(\rm{Tr}_A)$ yields the density matrix, $\rho  =
{\rm{T}}{{\rm{r}}_A}|{\Phi ^\rho }\rangle \langle {\Phi ^\rho }|.$
In other words, mixed states can always be seen as pure states of a
larger system (i.e., the system plus the introduced ancilla), the
topological invariant (called Chern value in
Ref.\cite{rivas13,viyuela86155140}) can then be  defined as
usual(closed system) TIs.

Turn to the topological invariant  for closed system  in more
details. The topological  invariant was first derived  by Thouless
{\it et al.}\cite{thouless82,kohmoto160343}, which provides a characterization of
fermionic time-reversal-broken (TRB) topological order in two
spatial dimensions. This was done by linear response theory in such
a way that the Hall conductivity is represented  in terms of a
topological invariant (or the Chern number), which  is  related to
an adiabatic change of the Hamiltonian in momentum space. However,
the extension of this topological invariant from closed to open
systems\cite{rivas13,viyuela86155140} is not given in this manner to date, i.e., it
is defined  neither via the Hall conductance, nor by the linear
response theory.

This paper presents a method  to extend the topological invariant
from closed   to open systems. The scheme is based on a linear
response theory developed here  for open systems. By calculating the
Hall conductance as a response to the adiabatic change of the
Hamiltonian in momentum space, the topological invariant  is
proportional to the quantized Hall conductivity for the system in
steady states.

\section*{Results}
To present the underlying principle of our method, we first extend
the Bloch's theorem to open system, then derive the Hall conductance
for  open systems.
\subsection*{Bloch's theorem and steady state.}
Take   isolated electrons in a potential as an example, the Bloch's
theorem for a closed system states that the energy eigenstate for an
electron in a periodic potential can be written as Bloch waves. To
extend this theorem from closed  to  open systems, we formulate this
statement as follows. Consider an electron in a periodic potential
$V(\vec{r})$ with periodicity $\vec{a}$, i.e.,
$V(\vec{r}+\vec{a})=V(\vec{r}).$ The one electron Schr\"odinger
equation
$$\left ( -\frac{\hbar^2}{2m}\nabla^2+V(\vec{r})\right
)\psi_n(\vec{r})=\varepsilon_n \psi_n(\vec{r})$$ should also have a
solution $\psi_n(\vec{r}+\vec{a})$ corresponding to the same energy
$\varepsilon_n$. Namely, $\psi_n(\vec{r}+\vec{a})=const\cdot
\psi_n(\vec{r})$. Here, $n$ denotes the index for the energy levels,
$m$ is the mass of electron. Furthermore, the energy eigenstate can
be written as,
\begin{equation}
\psi_n(\vec{r})=e^{i\vec{k}\cdot\vec{r}}
u_{n,\vec{k}}(\vec{r}),\label{cBT}
\end{equation}
where $u_{n,\vec{k}}(\vec{r})$ satisfies
$u_{n,\vec{k}}(\vec{r}+\vec{a})=u_{n,\vec{k}}(\vec{r})$ are the
Bloch waves, $\vec{k}$ denotes the Bloch vector.  Define a
translation operator $T_{\vec{a}}$ which, when operating on any
smooth function $f(\vec{r})$, shifts the argument by $\vec{a}$,
$T_{\vec{a}}f(\vec{r})=f(\vec{r}+\vec{a}).$ This operator can be
explicitly written as $T_{\vec{a}}=e^{i\vec{k}\cdot\vec{a}}.$ If
$T_{\vec{a}}$ is applied to a Hamiltonian $H=\left (
-\frac{\hbar^2}{2m}\nabla^2+V(\vec{r})\right )$ with periodic
potential $V(\vec{r})$, the Hamiltonian is left invariant, i.e.,
$[H,T_{\vec{a}}]=0.$

Now we extend the Bloch's theorem from closed to open systems.
Suppose that the density matrix $\rho$ of the open system is
governed  by a master equation \cite{gardiner04},
\begin{equation}
\dot \rho = - \frac{i}{\hbar}\left[ {H,\rho } \right] + {\cal L}
(\rho )\equiv {\cal P}(\rho), \label{mse}
\end{equation}
where ${\cal L} (\rho )$ sometimes called dissipator describes the
decoherence effect.  In the absence of decoherence, we know that a
key ingredient of the Bloch's theorem is $[H,T_{\vec{a}}]=0.$ Thus,
to preserve the translation invariant of the dynamics, it is natural
to restrict the master equation to satisfy
\begin{equation}
{\cal P}(T_{\vec{a}}\rho T_{\vec{a}}^{\dagger})=T_{\vec{a}}{\cal
P}(\rho) T_{\vec{a}}^{\dagger},\label{tio}
\end{equation}
which is similar to $H
T_{\vec{a}}|\psi\rangle=T_{\vec{a}}^{\dagger}H|\psi\rangle$ for a
closed system. For a Lindblad master equation with decay rates
$\gamma_j$ and Lindblad operators $F_j$\cite{gardiner04},
\begin{equation}
{\cal P}(\rho) = -\frac{i}{\hbar}[H,\rho] + \sum_j \gamma_j(2F_j\rho
F_j^\dagger- F_j^\dagger F_j \rho - \rho F_j^\dagger F_j),
\label{lbmse}
\end{equation}
Eq. (\ref{tio}) leads to $[F_j, T_{\vec{a}}]=0$ and
$[H,T_{\vec{a}}]=0$ for any $j$. Consequently, when $\rho_{ss}$ is a
steady state of the system,
$T_{\vec{a}}\rho_{ss}T_{\vec{a}}^{\dagger}$ is also a steady state,
since ${\cal P}(T_{\vec{a}}\rho_{ss}
T_{\vec{a}}^{\dagger})=T_{\vec{a}}{\cal P}(\rho_{ss})
T_{\vec{a}}^{\dagger}=0.$

The translation operator satisfying Eq. (\ref{tio}) preserve the
decoherence-free subspace(DFS)
\cite{zanardi97,lidar812594,shabani72042303,karasik77052301}. DFS
has been defined as a collection of states that undergo unitary
evolution in the presence of decoherence.  The theory of DFS
provides us with an important strategy to the passive presentation
of quantum information. The advantage of this
translation-preserved-DFS is its possible applications   into
quantum information processing in the presence of  decoherence.

Identifying the problem of energy eigenstates in closed system with
the problem of steady states in open system, we formulate the
Bloch's theorem of open system as follows. For an open system
described by Eq. (\ref{mse}) with translation invariant map ${\cal
P}$, its steady state can be written as\cite{rivas13,viyuela86155140},
\begin{equation}
\rho_{ss}=\sum_{m,n}\sum_{\vec{k}}\alpha_{mn}(\vec{k})
|u_{m,\vec{k}}\rangle \langle
u_{n,\vec{k}}|+\alpha_{00}|0\rangle\langle 0|, \label{oBT}
\end{equation}
where $\langle \vec{r}|u_{m,\vec{k}}\rangle=u_{m,\vec{k}}(\vec{r})$
are Bloch waves of the corresponding closed system, and $|0\rangle$
is the vacuum state. The coefficients $\alpha_{m,n}(\vec{k})$ are
independent of position $\vec{r}$, this fact can lift the limitation
on the uniqueness required  for steady states $\rho_{ss}$. In other
words, $T_{\vec{a}}\rho_{ss}T_{\vec{a}}^{\dagger}=\rho_{ss}$ satisfy
naturally in this situation.   For the Lindblad master equation Eq.
(\ref{lbmse}), $[T_{\vec{a}}, F_j]=0$ yields
$$T_{\vec{a}} F_j|u_{n,\vec{k}}\rangle=e^{i\vec{k}
\cdot \vec{a}}F_j|u_{n,\vec{k}}\rangle.$$ Thus, the Lindblad
operators $F_j$ conserve the crystalline momentum $\vec{k}$ of the
Bloch wave. This does not imply that the steady state has a
well-defined crystalline momentum, since the steady state is a
convex mixture of well-defined momenta states.

It is worth noticing that the Bloch's theorem of open system Eq.
(\ref{oBT}) relies on a postulate that the number of particles in
the system is limited to below 1. When the number of particles is
conserved,  and consider the system having only one particle, the
last term in Eq. (\ref{oBT}) can be omitted.

In the following, we shall restricted our attention to open systems
that possess translation invariance and preserve the TI phase. For
this purpose, we need to specify how the dissipator is realized in
physics. In an optical lattice setup, such a dissipative dynamics
can be engineered by manipulating  couplings of the lattice to
different atomic species, which  play  the role of the dissipative
bath\cite{verstraete09, diehl11, bardyn12, muller12, horstmann13,
eisert10}.

\subsection*{Linear response formula for the Hall conductance.}
To derive  the Hall conductance of an open system, we first develop
a perturbation theory to calculate  the steady state of the master
equation Eq.(\ref{mse}). Perturbation theory  is a widely accepted
tool in the investigation of closed quantum systems. In the context
of open quantum systems, however, the perturbation theory based on
the Markovian quantum master equation is barely  developed. The
recent investigation of open systems mostly relies on exact
diagonalization of the Liouville superoperator or quantum
trajectories, this approach  is limited by current computational
capabilities and is a drawback  for  analytically  understanding
open systems.

In a recent work\cite{yi00}, we have developed a perturbation theory
for open systems based on the Lindblad master equation. In this
approach, the decay rate was treated as a perturbation.  Successive
terms of those expansions yield characteristic loss rates for
dissipation processes. In Ref.\cite{valle13}, instead of computing
the full density matrix, the authors develop a perturbation theory
to  calculate directly the correlation functions. Based on the right
and left eigenstates of the superoperator $\cal P$, a perturbation
theory is proposed\cite{li13}, the non-positivity issue of the
steady-state may appear in this method due to  truncations. Here, we
apply the perturbation theory in Ref.\cite{yi00} to derive the
steady state. Instead of treating the decoherence as perturbation, a
perturbed term  in the Hamiltonian is introduced.

To present the main results of our method, we first consider a
situation without decoherence, namely, for an open system described
by the master equation, $$ \dot \rho = - \frac{i}{\hbar} [H,\rho] +
{\cal L} (\rho ),$$  we have $\langle \phi_i|{\cal
L}(\rho_{ss}^{(1)} )|\phi_j\rangle=0$, where $\rho_{ss}^{(1)}$ is
the first order expansion of steady state, $\rho_{ss}\simeq
\rho_{ss}^{(0)}+ \lambda\rho_{ss}^{(1)}=\sum\limits_{ij} {(\alpha
_{ij}^{(0)} + \lambda \alpha _{ij}^{(1)})} |{\phi _i}\rangle \langle
{\phi _j}|,$ $\lambda$ is the perturbation parameter from
$H=H_0+\lambda H^{\prime}$, $|{\phi_i}\rangle$ is an eigenstate of
$H_0$ with eigenvalue $\varepsilon_i$, $i$ is the index for the
eigenlevels. The steady state in this situation would be a diagonal
matrix in the basis of energy eigenstates due to thermalization,
i.e., $\alpha_{ij}^{(0)}=\alpha_{ii}^{(0)}\delta_{ij}$ with
$\delta_{ij},$ the Dirac delta function. The expansion coefficients
then reduce to,
\begin{equation}
-\alpha_{ij}^{(1)}(\varepsilon_i-\varepsilon_j)=
H_{ij}^{\prime}(\alpha_{jj}^{(0)}-\alpha_{ii}^{(0)}),
\end{equation}
obviously,
\begin{equation}
\alpha_{ij}^{(1)}=\frac{\alpha_{jj}^{(0)}
-\alpha_{ii}^{(0)}}{\varepsilon_j-\varepsilon_i}H_{ij}^{\prime},
\end{equation}
where $H_{ij}^{\prime}=\langle\phi_i|H^{\prime}|\phi_j\rangle.$  To
shorten the notation, here and hereafter, the perturbation parameter
$\lambda$ is included in $\alpha_{ij}^{(1)}$. Namely,
$\alpha_{ij}^{(1)}$ here and in the following equals the multiple of
$\alpha_{ij}^{(1)}$ and $\lambda$ in Eq. (\ref{vy}). Consider a
x-direction weak electric field, $\vec{E}=(E_x,0,0)$, simple algebra
yields(see Methods),
$$H_{mn}^{\prime}=ieE_x\langle\phi_m|\frac{\partial}{\partial
k_x}|\phi_n\rangle,$$ where $k_x$ is the x-component of $\vec{k}$,
$e$ is the charge of electron. Suppose the temperature is zero and
the single filled band is the s-th Bloch band, i.e., all
$\alpha_{ij}^{(0)}=0$ except $\alpha_{ss}^{(0)}=1$,
$\alpha_{st}^{(1)}$ takes ($t$ runs over the band indices),
$$\alpha_{st}^{(1)}=-\frac{ieE_x}{\varepsilon_t-\varepsilon_s}\langle\phi_s|\frac{\partial}
{\partial k_x}|\phi_t\rangle,$$  while $\alpha_{ij}^{(1)}=0$ for
other $i$ and $j$. Collecting all these results, we have
($\hbar=h/2\pi$, Planck constant)
$$\bar{v}_y=-i\frac{eE_x}{\hbar}\left [ \langle
\frac{\partial\phi_s}{\partial k_x}|\frac{\partial\phi_s}{\partial
k_y}\rangle-\langle\frac{\partial\phi_s}{\partial
k_y}|\frac{\partial\phi_s}{\partial k_x}\rangle\right ].$$ Here the
fact that the  contribution from the filled band is zero has been
used. This is exactly the results in\cite{thouless82,kohmoto160343,bohm03} for
closed systems.

Next let us consider what happens when there is a single steady band
in the presence of decoherence. We refer the single steady band to
that, with a fixed $\vec{k}$, there is only a single energy
eigenstate in the DFS. We denote this state by $|\phi_s\rangle$. In
this case, the operator $F_j$ in Eq. (\ref{lbmse}) may takes,
$F_j=|\phi_s\rangle\langle \phi_j|.$ This describes a situation
where all bands decay to the s-th band at  rates of $\gamma_j$ with
preserved momenta $\hbar\vec{k}$, see Fig.~1. Straightforward
calculation yields,
\begin{eqnarray}
\langle\phi_s|{\cal L}(\rho_{ss}^{(1)})|\phi_s\rangle&=&0,\nonumber\\
\langle\phi_s|{\cal
L}(\rho_{ss}^{(1)})|\phi_n\rangle&=&-\gamma_n\alpha_{sn}^{(1)},\ \
n\neq s,\nonumber\\
\langle\phi_m|{\cal
L}(\rho_{ss}^{(1)})|\phi_s\rangle&=&-\gamma_m\alpha_{ms}^{(1)}, \ \
m\neq s,\\ \label{dissM1} \langle\phi_m|{\cal
L}(\rho_{ss}^{(1)})|\phi_n\rangle&=&-(\gamma_m+\gamma_n)
\alpha_{mn}^{(1)},\ \  m, n\neq s.\nonumber
\end{eqnarray}
\begin{figure}[h]
\centering
\includegraphics[scale=0.5]{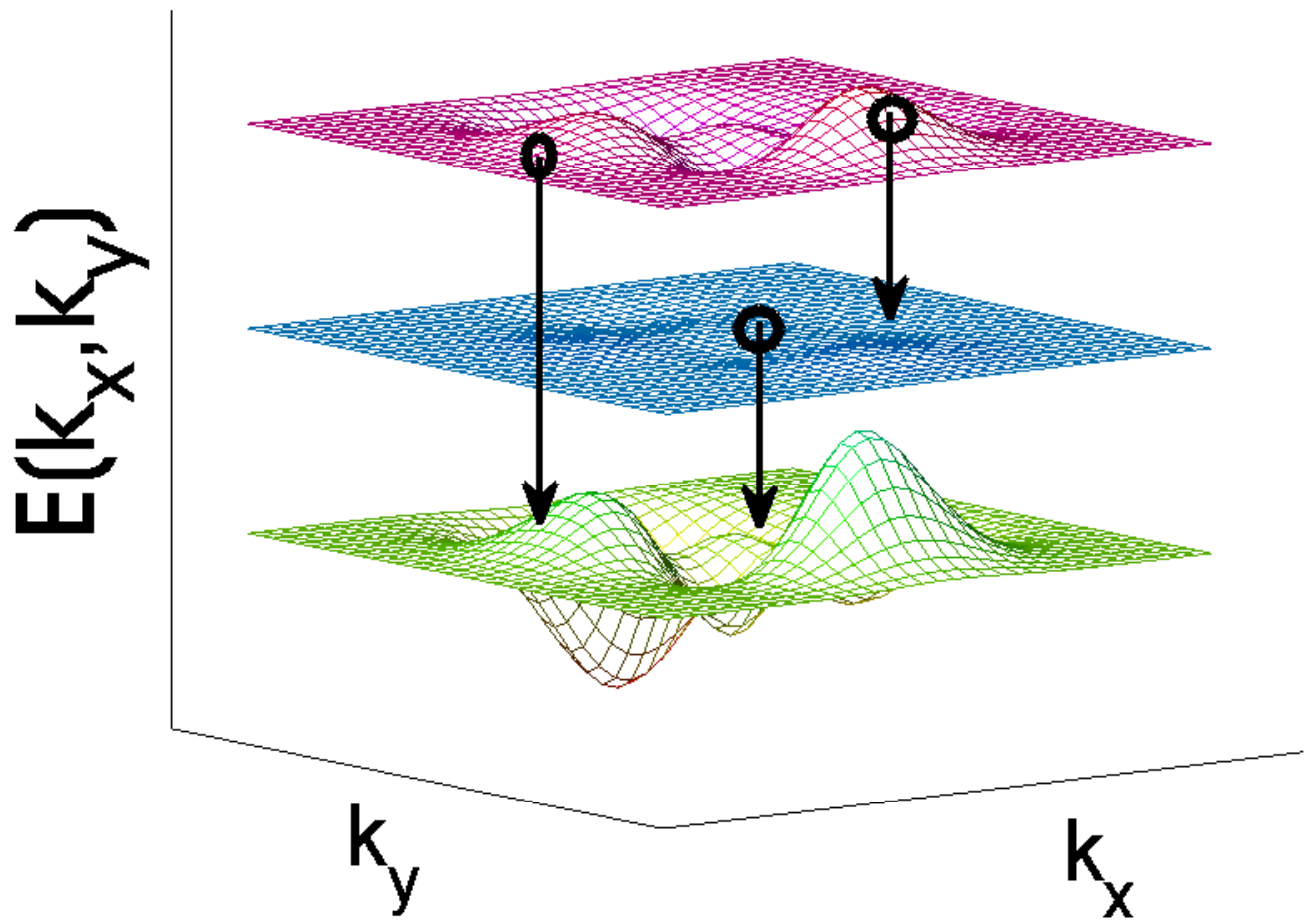}
\caption{(Color online)Illustration of the  decoherence
mechanism--decays from upper bands to the lowers.} \label{1}
\end{figure}
Substituting these equations into Eq. (\ref{1thalpha}) and using
$\alpha_{ij}^{(0)}=0$ for any $i$ and $j$ except
$\alpha_{ss}^{(0)}=1,$ we arrive at
\begin{equation}
\alpha_{sn}^{(1)}= -ieE_x\frac{\langle
\phi_s|\frac{\partial}{\partial
k_x}|\phi_n\rangle}{\varepsilon_n-\varepsilon_s+i\Delta_{sn}}.\label{alphamn}
\end{equation}
Here $\Delta_{sn}$ is defined as $\Delta_{sn}=
\gamma_n\cdot(1-\delta_{ns}),$ and $\alpha_{mn}^{(1)}=0$ for $m\neq
s$ and $n\neq s$. For large energy band gaps,
$|\varepsilon_m-\varepsilon_n|\gg (\gamma_m+\gamma_n),$ the
coefficients approximately take,
\begin{equation}
\alpha_{sn}^{(1)}\simeq -ieE_x\frac{\langle
\phi_s|\frac{\partial}{\partial
k_x}|\phi_n\rangle}{\varepsilon_n-\varepsilon_s}\left(1-i\frac{\Delta_{sn}}
{\varepsilon_n-\varepsilon_s}-(\frac{\Delta_{sn}}
{\varepsilon_s-\varepsilon_n})^2\right ).\label{alphamnA}
\end{equation}

It is not trivial to extend the case of single steady band to two
 steady bands, as we shall show below. Denote the two steady bands by
$|\phi_{s_1}\rangle$ and $|\phi_{s_2}\rangle$, respectively, a
possible realization of the two steady bands is via a dissipator,
\begin{equation}
{\cal L}(\rho) = \sum_{\substack{\alpha=1,2 \\ j=1,...,N}}
\gamma_{\alpha j}(2F_{\alpha j}\rho F_{\alpha j}^\dagger- F_{\alpha
j}^\dagger F_{\alpha j} \rho - \rho F_{\alpha j}^\dagger F_{\alpha
j}), \label{lbmse2}
\end{equation}
where we choose $F_{\alpha j}=|\phi_{s_{\alpha}}\rangle\langle
\phi_j|,$ and $\gamma_{\alpha j}$ denotes the decay rate.  Following
the same procedure as in the case of single steady band, we find
$\alpha_{mn}^{(1)}$ can be written in a form similar to Eq.
(\ref{alphamnA}),
\begin{eqnarray}
\alpha_{s_{\alpha}n}^{(1)}&\simeq &-ieE_x\frac{\langle
\phi_{s_{\alpha}}|\frac{\partial}{\partial
k_x}|\phi_n\rangle}{\varepsilon_n-\varepsilon_{s_{\alpha}}}\nonumber\\
&\cdot& \left(1-i\frac{\Delta_{s_{\alpha}n}}
{\varepsilon_n-\varepsilon_{s_{\alpha}}}-(\frac{\Delta_{s_{\alpha}n}}
{\varepsilon_{s_{\alpha}}-\varepsilon_n})^2\right )
 \alpha_{s_{\alpha}s_{\alpha}}^{(0)},\nonumber\\
\alpha_{s_1s_2}^{(1)}&\simeq & ieE_x\frac{\langle
\phi_{s_1}|\frac{\partial}{\partial
k_x}|\phi_{s_2}\rangle}{\varepsilon_{s_2}
-\varepsilon_{s_1}}(2\alpha_{s_2s_2}^{(0)}-1). \label{alphamn2}
\end{eqnarray}
with  $\Delta_{s_{\alpha}n}$ defined by,
\begin{eqnarray}
\Delta_{s_{\alpha}n}=
(\gamma_{s_1n}+\gamma_{s_2n})\cdot(1-\delta_{ns}),
\end{eqnarray}
where $\delta_{ms}=1$ when $s=s_1$ or $s_2$, otherwise it takes 0.
Substituting $\alpha_{mn}^{(1)}$ into the Hall current and supposing
the current  is zero in the absence of the external field, we find
that the Hall current can be separated into two parts. The first
part is independent of the decay rates and it can be written in
terms of Chern number, while the second part takes a different form
related closely to the dissipator. These two parts also manifest in
the Hall conductivity discussed below, suggesting us to define a
{\it topological} value called Chern rate for the system.

The Hall conductivity, defined as the ratio of the Hall current
density $j_H$ and the electronic field $E_x$, is therefore given by
$\sigma_H=\sigma_H^{(0)}+\delta\sigma_H=\sigma_H^{(0)}+\delta\sigma_H^{(1)}
+\delta\sigma_H^{(2)}.$ Here
\begin{widetext}
\begin{eqnarray}
\sigma_H^{(0)}&=&\frac{e^2}{h}\int\frac{idk_xdk_y}{2\pi}
\sum_{\alpha=1,2}\alpha_{s_{\alpha}s_{\alpha}}^{(0)} \left( \langle
\frac{\partial\phi_{s_{\alpha}}}{\partial
k_x}|\frac{\partial\phi_{s_{\alpha}}}{\partial
k_y}\rangle-\langle\frac{\partial\phi_{s_{\alpha}}}{\partial
k_y}|\frac{\partial\phi_{s_{\alpha}}}{\partial k_x}\rangle\right
),\\ \label{0thhall}
\delta\sigma_H^{(1)}&=&\frac{e^2}{h}\int\frac{dk_xdk_y}{2\pi}\sum_{\alpha=1,2}\sum_{j\neq
s_{\alpha}}\alpha_{s_{\alpha}s_{\alpha}}^{(0)}
\frac{\Delta_{js_{\alpha}}}{\varepsilon_j-\varepsilon_{s_{\alpha}}}\left
(\langle \frac{\partial\phi_{s_{\alpha}}}{\partial
k_x}|\phi_j\rangle\langle
\phi_j|\frac{\partial\phi_{s_{\alpha}}}{\partial k_y}\rangle
+\langle \frac{\partial\phi_{s_{\alpha}}}{\partial
k_y}|\phi_j\rangle\langle
\phi_j|\frac{\partial\phi_{s_{\alpha}}}{\partial k_x}\rangle\right ), \nonumber\\
\delta\sigma_H^{(2)}&=&-\frac{e^2}{h}\int\frac{idk_xdk_y}{2\pi}\sum_{\alpha=1,2}\sum_{j\neq
s_{\alpha}}\alpha_{s_{\alpha}s_{\alpha}}^{(0)} \left (
\frac{\Delta_{js_{\alpha}}}{\varepsilon_j-\varepsilon_{s_{\alpha}}}\right
)^2\left (\langle \frac{\partial\phi_{s_{\alpha}}}{\partial
k_x}|\phi_j\rangle\langle
\phi_j|\frac{\partial\phi_{s_{\alpha}}}{\partial k_y}\rangle
-\langle \frac{\partial\phi_{s_{\alpha}}}{\partial
k_y}|\phi_j\rangle\langle
\phi_j|\frac{\partial\phi_{s_{\alpha}}}{\partial k_x}\rangle\right)
\label{1thhall}
\end{eqnarray}
\end{widetext}
 To derive these results, $\langle
\phi_i|\frac{\partial H_0}{\partial k}|\phi_j\rangle
=(\varepsilon_i-\varepsilon_j)\langle \frac{\partial
\phi_i}{\partial k}|\phi_j\rangle$ has been used. This is one of the
main result of this work. It is worth pointing out that this
result sharply depends on the decoherence mechanism. In fact, as we
will  show later in the two-band model, the Hall conductivity is not
a mixture of Hall conductivities for various steady bands.

Assume  $\alpha^{(0)}_{s_{\alpha}s_{\alpha}}$ independent of $k_x$
and $k_y$, the integral on the right hand side of $\sigma_H^{(0)}$,
i.e.,
\begin{equation}
C_{s_{\alpha}}\equiv \int\frac{idk_xdk_y}{2\pi}\left( \langle
\frac{\partial\phi_{s_{\alpha}}}{\partial
k_x}|\frac{\partial\phi_{s_{\alpha}}}{\partial
k_y}\rangle-\langle\frac{\partial\phi_{s_{\alpha}}}{\partial
k_y}|\frac{\partial\phi_{s_{\alpha}}}{\partial k_x}\rangle\right ),
\end{equation}
is nothing  but the Chern number which takes integer values as
pointed out in\cite{bohm03}. Then $\sigma_H^{(0)}$ can be written
as
$$\sigma_H^{(0)}=\frac{e^2}{h}
\sum_{\alpha=1,2}\alpha^{(0)}_{s_{\alpha}s_{\alpha}}C_{s_{\alpha}}.$$
$\sigma_H^{(0)}$ is a weighted Chern number for the two steady
bands. This term may not be an integer for a general open system,
despite its topological origin. For  $\vec{k}$-dependent
$\alpha^{(0)}_{s_{\alpha}s_{\alpha}}$, $\int {\frac{{id{k_x}d{k_y}}}{{2\pi }}} \alpha _{{s_\alpha }{s_\alpha }}^{(0)}[\langle \frac{{\partial {\phi _{{s_\alpha }}}}}{{\partial {k_x}}}|\frac{{\partial {\phi _{{s_\alpha }}}}}{{\partial {k_y}}}\rangle  - \langle \frac{{\partial {\phi _{{s_\alpha }}}}}{{\partial {k_y}}}|\frac{{\partial {\phi _{{s_\alpha }}}}}{{\partial {k_x}}}\rangle ]$
has been defined as the so-called Chern value\cite{rivas13,viyuela86155140}, which
witnesses  a topological non-trivial order present in the Berry
curvature. It recovers the standard Chern number if the steady state
is a pure Bloch state.

$\delta\sigma_H$ consists of two parts,
$\delta\sigma_H=\delta\sigma_H^{(1)} +\delta\sigma_H^{(2)}.$ Here,
$\delta\sigma_H^{(1)}$ and  $\delta\sigma_H^{(2)}$ describe
respectively the first order and second order corrections of the
decoherence to the Hall conductivity. They can not be written in
terms of Chern number in general, since both $\Delta_{mn}$ and the
energy gap depend on band index. Therefore, there is no topological
invariance for the open system from the viewpoint of Hall
conductivity, this is true even when the dissipation rates
$\gamma_j$ and the band gaps are independent of band index,
$\delta\sigma_H^{(2)}$  can be expressed in terms of Chern numbers
in this case, but $\delta\sigma_H^{(1)}$ still can  not. The Hall
current given by $\delta\sigma_H$ characterizes the environmental
activation of excited electrons in the bulk,  and it is not zero in
the regions outside the topological regime, where
$C_{s_{\alpha}}=0$. This can be found in Eq. (\ref{1thhall}).

These observations motivate us to define a topological value, to
which we will refer as Chern rate,
\begin{equation}
C^{\rho}\equiv\sigma_H\frac{h}{e^2}.
\end{equation}
We adopt terminology Chern rate for the following reasons. Firstly,
it possesses topological origin; Secondly, it may not take an
integer for a general open system; Thirdly, it should differ from
the Chern value defined in Ref.\cite{rivas13,viyuela86155140}, and in addition the
Hall conductance is simply a multiple of the Chern rate and
$\frac{e^2}{h}$. Of cause, the Chern rate returns back to the Chern
number when the system is an isolated topological insulator. It is
well known that Bloch's waves $u_{n,\vec k}(r)$ under time-reversal
transformation take ${\cal T} u_{n,\vec k}(r)=u^*_{n,-\vec k}(r)$,
then the Berry curvature defined by $F_n(\vec k)=i\left( \langle
\frac{\partial u_{n,\vec k}}{\partial k_x}|\frac{\partial u_{n,\vec
k}}{\partial k_y}\rangle-\langle\frac{\partial u_{n,\vec
k}}{\partial k_y}|\frac{\partial u_{n,\vec k}}{\partial
k_x}\rangle\right )$ under the time-reversal transformation
satisfies, ${\cal T} F_n(\vec k)=-F_n(-\vec k)$. So, for system with
time reversal symmetry, $F_n(\vec k)$ is an odd function of $\vec
k$. As a consequence, the Chern number for a time-reversal invariant
system is zero, because the integral of an odd function over the
whole Brillouin zone must be zero. This is not the case for second
line in Eq. (\ref{1thhall}) that is an even function of $k$. This
fact reflects that the second line in Eq. (\ref{1thhall}) may not be
zero for a time-reversally    invariant system, and hence the Chern
rate loses  partially  its topological origin in this case. We will
illustrate below that this non-topological term can be eliminated by
properly designing $\varepsilon$ and $\Delta$ in Eq. (\ref{H01}).

We now  apply this formalism to derive a formula for Hall
conductance in a two-band system. A decoherence mechanism
different from this section is considered, namely the decoherence
operator $F_j$ in the dissipator is not purely a Jordan block. This
difference would manifest in the Hall conductivity, for example, the
Hall conductivity is not a mixture of Hall conductivities for
various bands.

\subsection*{Applications of the formalism to a two-band model.}
We can apply the representation  to develop a general formula for
Hall conductance for a two-band system. Let us start with an
effective Hamiltonian,
\begin{eqnarray}
H &=& \sum\limits_{k_x^0,{k_y}} {h_S(k_x^0,{k_y}),} \nonumber\\
h_S(k_x^0,{k_y}) &=& \epsilon(k)+\left(\begin{array}{cc} d_z &
d_x-id_y\\
d_x+id_y & -d_z \end{array}\right )\nonumber\\
&\equiv& \epsilon(k)+\left( {\begin{array}{*{20}{c}}
\varepsilon &{\Delta {e^{ - i\varphi}}}\nonumber\\
{\Delta {e^{i\varphi}}}&{ - \varepsilon }
\end{array}} \right), \nonumber\\ \label{H01}
\end{eqnarray}
where $\epsilon$ is the  energy without couplings, it may take
$\frac{\hbar^2k^2}{2m^*}$ for the band electron with effective mass
$m^*$, and $(E_0-Dk^2)$ with constant $E_0$ and $D$ for the surface
states of bulk $\rm{Bi_2Se_3}$\cite{lu10}. $d_j=d_j(k_x^0,k_y)$ are
the momentum-dependent coefficients which describe the spin-orbit
couplings. $\varepsilon=d_z$, $\Delta=\sqrt{d_x^2+d_y^2},$
$\tan\varphi=\frac{d_y}{d_x},$ and  $k^2=(k_x^0)^2+k_y^2$.

Consider phenomenally a dissipator,
\begin{equation}
{\cal L} (\rho ) = \sum\limits_{k_x^0,{k_y}} \gamma(2{\sigma _ -
}\rho {\sigma _ + } - {\sigma _ + }{\sigma _ - }\rho  - \rho {\sigma
_ + }{\sigma _ - }),
\end{equation}
where $\gamma=\gamma(k_x^0,{k_y})$ are momentum dependent decay
rates, $\sigma_j=\sigma_j(k_x^0,k_y),\ (j=+,-)$ are Pauli matrices.
This dissipator describes a decay of the fermion from the spin-up
state to the spin-down state with conserved momenta. It
differs from those in the last section at that this dissipator  does
not describe decays from one band to the other,  it instead
characterizes  the decay of the electron spin states, see Fig.~2.

\begin{figure}[h]
\centering
\includegraphics[scale=0.5]{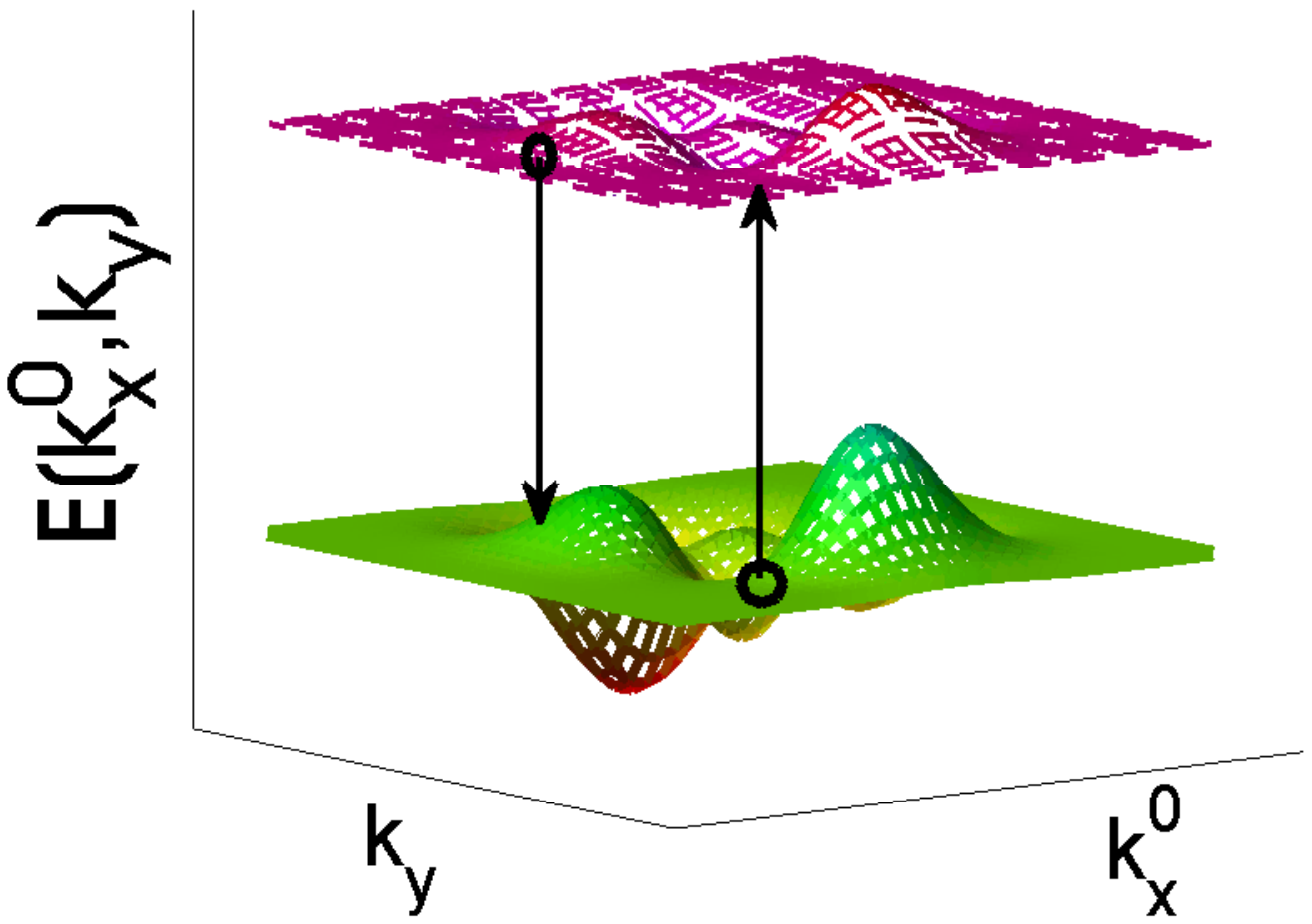}
\caption{(Color online)Illustration of the
decoherence mechanism. It not only leads to a decay from the upper
band to the low band but also a flip from the lower to the upper.
Besides, it induces dephasings for each bands.} \label{2}
\end{figure}

Now we introduce a perturbation $\lambda h^{\prime}$ to Hamiltonian
$h_S(k_x^0,{k_y})$, the total Hamiltonian with fixed $k_x^0$ and
$k_y$ is then $h_S(k_x^0,{k_y})+\lambda h^{\prime}$. Up to first
order in $\lambda$, we write the steady state with fixed $k_x^0$ and
$k_y$ as, $\tau = {\tau^{(0)}} + \lambda \tau^{(1)}.$ Tedious but
straightforward calculations yield,
\begin{equation}
\tau^{(1)} = \left( {\begin{array}{*{20}{c}}
\tau^{(1)}_{11}&\tau^{(1)}_{12}\\
{{\tau^{(1)}_{21}}}&{ - {\tau^{(1)}_{11}}}
\end{array}} \right),
\label{fs}
\end{equation}
in the basis spanned by the eigenstates of $h_S(k_x^0,{k_y})$, we
have
\begin{eqnarray}
{\tau^{(1)}_{12}} &=& \frac{{{s_1} - {s_2} +
2{\rm{i}}{s_3}({h^{\prime}_{11}} - {h^{\prime}_{22}})}}{{4{{[{\gamma
^2}
+ 3{E_1 ^2} + {E_1 ^2}\cos(2\theta )]}^2}}},\nonumber\\
{\tau^{(1)}_{21}} &=& \tau^{(1)*}_{12} , \label{coefficient}
\end{eqnarray}
and
\begin{eqnarray}
 s_1 &=& \cos\theta (\gamma -2iE_1) [-
4i\gamma^2h^{\prime}_{12} +  \gamma E_1(h^{\prime}_{12} +
h^{\prime}_{21}) \nonumber\\
&&- 14iE_1^2h^{\prime}_{12}],\nonumber\\
s_2 &=& E_1 \cos(3\theta )(\gamma-2iE_1)[\gamma
(h^{\prime}_{12} + h^{\prime}_{21})+ 2iE_1 h^{\prime}_{12}],\\
s_3&=& \gamma \sin\theta [-\gamma^2 + 3i\gamma E_1+ 3E_1^2+E_1
\cos(2\theta )(i\gamma +E_1 )].\nonumber \label{coefficient10}
\end{eqnarray}
Here, $E_1=\sqrt{\varepsilon^2+\Delta^2}$,
$\cos\theta=\frac{\varepsilon}{\sqrt{\varepsilon^2+\Delta^2}}$,
$h_{ij}^{\prime},i,j=1,2$ are matrix elements of $h^{\prime}$ in the
basis spanned by the eigenstates of $h_S(k_x^0,k_y)$. For more
details, see Methods. The diagonal elements of $\tau^{(1)}$ is not
listed here, since it has no contribution to the conductivity.  In
weak dissipation limit, $\gamma\rightarrow 0,$ we can expand
$\tau^{(1)}_{12}$ in powers of $\gamma$.  To first order in
$\gamma$, $\tau^{(1)}_{12}$ can be written as,
\begin{eqnarray}
\tau _{12}^{(1)} &\simeq&  - \frac{{7\cos \theta  + \cos 3\theta }}{{{E_1}{{(3 + \cos 2\theta )}^2}}}h_{12}^\prime  - \frac{{i\gamma (7\cos \theta  + \cos 3\theta )h_{12}^\prime }}{{2E_1^2{{(3 + \cos 2\theta )}^2}}} \nonumber\\
 &&+ \frac{{i\gamma (\cos 3\theta  - \cos \theta )(h_{12}^\prime  + h_{21}^\prime )}}{{2E_1^2{{(3 + \cos 2\theta )}^2}}}\nonumber\\
& &+ \frac{{i\gamma (3\sin \theta  + \sin \theta \cos 2\theta )(h_{11}^\prime  - h_{22}^\prime )}}{{2E_1^2{{(3 + \cos 2\theta )}^2}}}.
\end{eqnarray}
Assuming a weak electric field is applied along the x-direction and
the corresponding vector potential is time-dependent, we find by
simple algebra that,  $h_{mn}^{\prime}=i\langle
\Phi_m|\frac{\partial}{\partial k_x^0}|\Phi_n\rangle E_xe.$
Substituting these equations into the Hall conductivity and assuming
$\gamma$ independent of $\vec{k}$, we have
\begin{widetext}
\begin{eqnarray}
\sigma_H&=&\frac{e^2}{h}\int\frac{idk_x^0dk_y}{2\pi}
\frac{2\cos\theta}{1+\cos^2\theta} \left( \langle
\frac{\partial\Phi_{E_2}}{\partial
k_x^0}|\frac{\partial\Phi_{E_2}}{\partial
k_y}\rangle-\langle\frac{\partial\Phi_{E_2}}{\partial
k_y}|\frac{\partial\Phi_{E_2}}{\partial k_x^0}\rangle\right
)\nonumber\\
&+&\frac{\gamma e^2}{h}\int\frac{dk_x^0dk_y}{2\pi}
\frac{2\cos\theta}{ E_1(1+\cos^2\theta)^2} \left( \langle
\frac{\partial\Phi_{E_2}}{\partial
k_x^0}|\Phi_{E_1}\rangle\langle\Phi_{E_1}|\frac{\partial\Phi_{E_2}}{\partial
k_y}\rangle+\langle\frac{\partial\Phi_{E_2}}{\partial
k_y}|\Phi_{E_1}\rangle\langle\Phi_{E_1}|\frac{\partial\Phi_{E_2}}{\partial
k_x^0}\rangle\right
)\nonumber\\
&-&\frac{\gamma e^2}{h}\int\frac{dk_x^0dk_y}{2\pi} \frac{
\sin^2\theta\cos\theta}{ E_1 (1+\cos^2\theta)^2} \left( \langle
\frac{\partial\Phi_{E_2}}{\partial
k_y}|\Phi_{E_1}\rangle\langle\frac{\partial\Phi_{E_2}}{\partial
k_x^0}|\Phi_{E_1}\rangle+\langle \frac{\partial\Phi_{E_1}}{\partial
k_y}|\Phi_{E_2}\rangle\langle\frac{\partial\Phi_{E_1}}{\partial
k_x^0}|\Phi_{E_2}\rangle\right
).\nonumber\\
\end{eqnarray}
\end{widetext}
Discussions on the Hall conductivity are in order. The first
integral describes a contribution of zeroth order in $\gamma$.
It is different from the usual Hall conductivity of TIs with
a single filled band $|\Phi_2\rangle$, the difference comes from the
deviation of the steady state from the Gibbs states. Note that when
$\Delta=0$, the first integral represents  the usual Hall
conductivity,  the second and third integral represent a correction
of dissipation to the Hall conductivity. We observe that the third
integral vanishes with $\Delta=0$. In this case, the second integral
reduces to,
\begin{eqnarray}
&&\frac{{\gamma {e^2}}}{{2{E_1}h}}\int {\frac{{dk_x^0d{k_y}}}{{2\pi }}} [\langle \frac{{\partial {\Phi _{{E_2}}}}}{{\partial k_x^0}}|{\Phi _{{E_1}}}\rangle \langle {\Phi _{{E_1}}}|\frac{{\partial {\Phi _{{E_2}}}}}{{\partial {k_y}}}\rangle \nonumber \\
&&+ \langle \frac{{\partial {\Phi _{{E_2}}}}}{{\partial {k_y}}}|{\Phi _{{E_1}}}\rangle \langle {\Phi _{{E_1}}}|\frac{{\partial {\Phi _{{E_2}}}}}{{\partial k_x^0}}\rangle ] \nonumber
\end{eqnarray}
which is exactly the result in the last section for  TIs with  two
bands. Noting that $|\Phi_{E_1}\rangle$ and $|\Phi_{E_2}\rangle$ can
be written in terms of $\theta$ and $\varphi$, $|\Phi_{E_1}\rangle=
\left(
\begin{array}{c}
\cos\frac{\theta}{2}e^{-i\varphi}\\
\sin\frac{\theta}{2}
\end{array} \right)$, $|\Phi_{E_2}\rangle=
\left( \begin{array}{c}
-\sin\frac{\theta}{2}e^{-i\varphi}\\
\cos\frac{\theta}{2}
\end{array} \right).$ We deduce the Hall conductance as,
$\sigma_H=\sigma_H^{(0)}+\delta\sigma_H^{(1)}$,
\begin{equation}
\begin{aligned}
\sigma _H^{(0)}=& \frac{{{e^2}}}{h}\int {\frac{{dk_x^0d{k_y}}}{{2\pi }}} \frac{{\sin 2\theta }}{{3 + \cos 2\theta }}\left( {\frac{{\partial \theta }}{{\partial k_x^0}}\frac{{\partial \varphi }}{{\partial {k_y}}} - \frac{{\partial \theta }}{{\partial {k_y}}}\frac{{\partial \varphi }}{{\partial k_x^0}}} \right),\\
\delta \sigma _H^{(1)}= &\gamma \frac{{{e^2}}}{h}\int {\frac{{dk_x^0d{k_y}}}{{2\pi }}} [\frac{{\cos \theta }}{{2{E_1}(1 + {{\cos }^2}\theta )}}\frac{{\partial \theta }}{{\partial k_x^0}}\frac{{\partial \theta }}{{\partial {k_y}}}\\
 &+ \frac{{\cos \theta {{\sin }^2}\theta (1 + 0.5{{\sin }^2}\theta )}}{{{E_1}{{(1 + {{\cos }^2}\theta )}^2}}}\frac{{\partial \varphi }}{{\partial k_x^0}}\frac{{\partial \varphi }}{{\partial {k_y}}}].
\label{shenhallf}
\end{aligned}
\end{equation}
This equation is available for all two-band system described by the
Hamiltonian in Eq. (\ref{H01}).

To be specific, we consider a two-dimensional ferromagnetic electron
gas with both Rashba and Dresselhaus coupling, this system can be
described by Hamiltonian Eq. (\ref{H01}) with $d_x=\lambda p_y-\beta
p_x$, $d_y=-\lambda p_x-\beta p_y$, and $d_z=h_0$, here  the momenta
$p_x=\hbar k^0_x$ and $p_y=\hbar k_y$. Using the formula Eq.
(\ref{shenhallf}) for the Hall conductivity, we calculate the Hall
conductivity and show the numerical results in Fig.~3.  Fig.~3(a)
shows the zero-order Hall conductivity versus $\beta$. The red-solid
line is for the closed system, while the blue-dashed line for the
open system with $\gamma\rightarrow 0$. It is interesting to notice
that Hall conductivity of the  open system with $\gamma\rightarrow
0$ is different from that in closed system. This is easy to
understand, the steady state of an open system is in general a mixed
state, even though the decoherence rate is close to zero. Fig.~3(a)
shows a phase transition at $\beta=\beta_c=\lambda$,  when
$\beta<\beta_c$, the Chern number of the closed system is 1, while
for $\beta>\beta_c$, the Chern number is -1. For open system, the
phase transition can still be found from the Hall conductivity, even
if the absolute value of  $\delta \sigma _H^{(0)}$ in the open
system is smaller than that in the closed system. The first-order
correction $\delta \sigma _H^{(1)}$ are negative on both sides of
$\beta_c$, as shown in Fig.~3(b), where we plot the first-order Hall
conductivity as a function of $\beta$ and $h_0$.

\begin{figure}[h]
\centering
\includegraphics[scale=0.4]{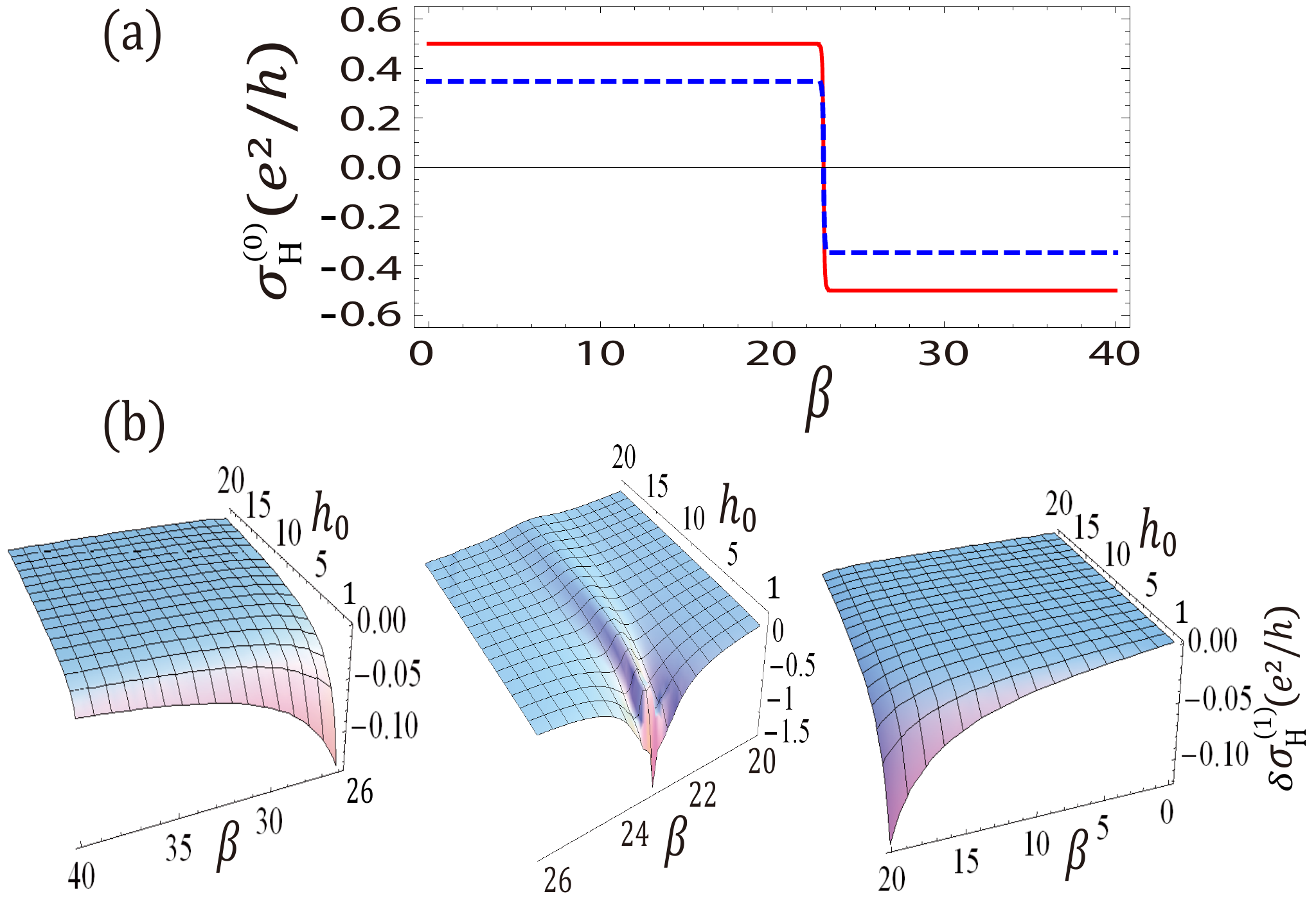}
\caption{(Color online)The zero-order and first-order conductivity
$\sigma _H^{(0)}$ and $\delta \sigma _H^{(1)}$ as a function of
$\beta$ (meV$ \cdot $nm/$\hbar$) and $h_0$ (in units of meV).
 Parameters chosen are, (a) $\gamma\rightarrow 0$, $\lambda=23$
meV$\cdot $nm/$\hbar$,  and (b)$\gamma$=0.1 meV, $\lambda=23$ meV$
\cdot $nm/$\hbar$. Note that $\sigma _H^{(0)}$ is independent of
$h_0$.} \label{3}
\end{figure}
The second concrete example  is bulk ${\rm Bi_2Se_3}$. The low-lying
effective model for bulk ${\rm Bi_2Se_3}$ can be formally
diagonalized, which can be interpreted  as the $K$ and $K^{\prime}$
valleys in the graphene\cite{lu10}. For the valleys located at $K$,
the effective Hamiltonian takes the same form as in Eq. (\ref{H01})
but with $d_x=\hbar v_F k_y$, $d_y=-\hbar v_F k_x$, and
$d_z=(\frac{\Delta_0}{2}-Bk^2)$. A straightforward calculation shows
that the term proportional to $\gamma$ in the Hall conductivity is
zero, this does not mean that the decoherence has no effect on the
Hall conductivity. In fact, the decoherence leads the system to a
mixed state, yielding the Hall conductivity,
\begin{equation}
\sigma_H=\frac{e^2}{2h}{\rm ln} \frac{1+{\rm sgn}^2B}{1+{\rm
sgn}^2\Delta_0}.
\end{equation}
For $B\neq 0$ and $\Delta_0\neq 0$, the Hall conductance is zero. For
$B= 0$ and $\Delta_0\neq 0$, $\sigma_H=-\frac{e^2}{2h}{\rm ln}2$, and
$\sigma_H=\frac{e^2}{2h}{\rm ln}2$ when $B\neq 0$ and $\Delta_0= 0$.
This is different from the results of closed system\cite{lu10}.

In the third concrete example, we apply the Hamiltonian Eq.
(\ref{H01}) to model the two-dimensional lattice in a magnetic field
\cite{kohmoto89}. The tight-binding Hamiltonian for such a lattice
is written as,
\begin{equation}
H=-t_a\sum\limits_{\left\langle {i,j} \right\rangle } {_x}
c_j^\dagger {c_i}{e^{i{\theta _{ij}}}}-t_b\sum\limits_{\left\langle
{i,j} \right\rangle } {_y} c_j^\dagger {c_i}{e^{i{\theta _{ij}}}},
\end{equation}
where $c_j$ is the usual fermion operator on the lattice, $t_a$ and
$t_b$ denote the hopping amplitudes along the x- and y-direction,
respectively. The first summation is taken over all the
nearest-neighbor sites along the x-direction and the second sum
along the y-direction. The phase $\theta_{ij}=-\theta_{ji}$
represents the magnetic flux through the lattice.  When $t_b=0$, the
single band $E(k_x)$ is doubly degenerate. The term with $t_b$ in
the Hamiltonian gives the coupling between the two branches of the
dispersion.  Consider two branches which are coupled by $|l|-$th
order perturbation, the gaps open and the size of the gap due to
this coupling is the order of $t_{b}^{|l|}$. The effective
Hamiltonian then take Eq. (\ref{H01})\cite{kohmoto89} with
$\varphi=k_y l$, $\varepsilon  = 2{t_a}\cos (k_x^0 + 2\pi
\frac{q}{p}m)$, and $\delta$ is proportional to (is the order of)
$t_b^{\left| l \right|}.$ In terms of $d_x$, $d_y$ and $d_z$, the
model takes, $d_x=\delta \cos(k_y l)$, $d_y=\delta \sin(k_y l)$, and
$d_z= 2t_a \cos(k_x+2\pi \frac{p}{q}m)$.

\begin{figure}[h]
\centering
\includegraphics[scale=0.4]{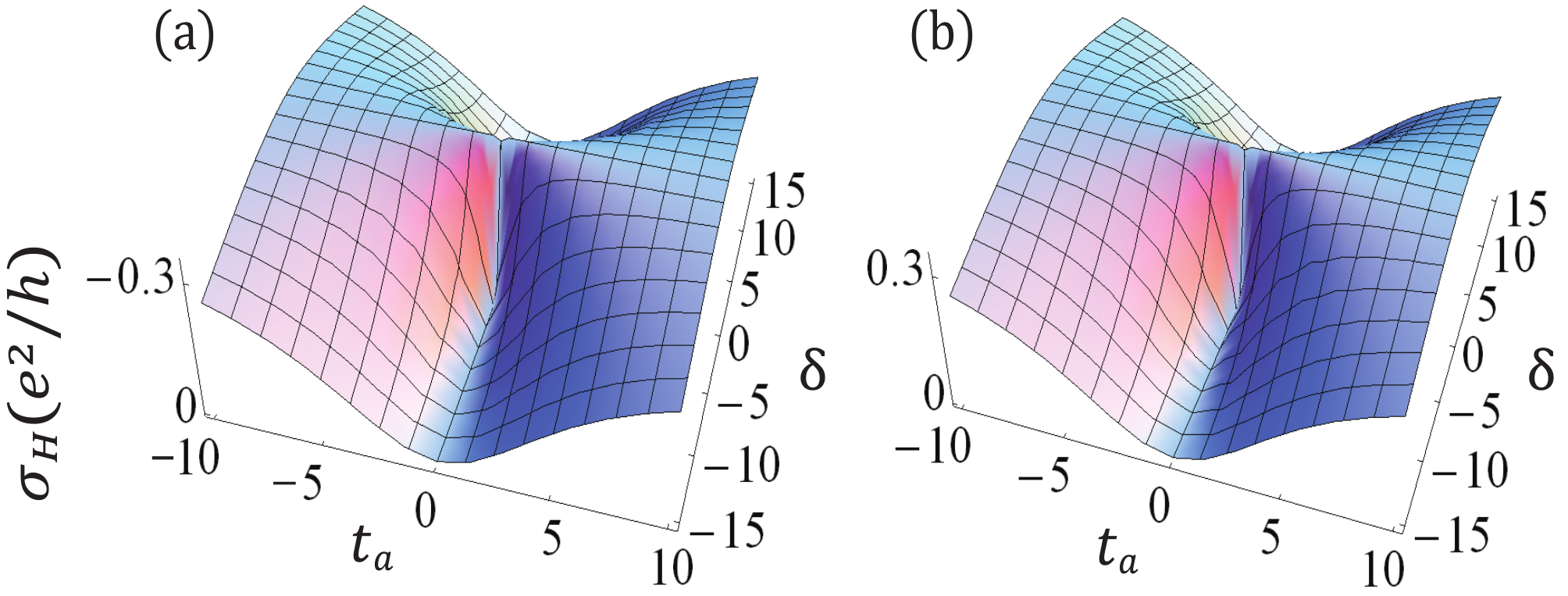}
\caption{(Color online)The conductivity $\sigma_H$ as a function of
$\delta$ (in units of meV) and $t_a$ (in units of meV). Parameters
chosen are $p = 1, q = 4, l = 1,$  (a) $m = 1$, and (b) $m = 2$.
Note that the sign of $\sigma_H$ in figures (a) and (b) are
different. Further numerical simulations show that $\sigma_H$
depends only on  the parity of $m$, i.e., figure (a) is   for all
odd $m$, while figure (b) for even $m$.} \label{4}
\end{figure}

When applying the formula to this model,  we can prove that
$h^{\prime}_{12}+h^{\prime}_{21}=0$ and
$h^{\prime}_{22}=h^{\prime}_{11}=0$. This can be done  by examining
the definition,
$h^{\prime}_{ij}=-i\hbar\langle\Phi_{E_i}|\frac{\partial}{\partial
t}|\Phi_{E_j}\rangle,$ and replacing  $k_x^0$ in Eq. (\ref{H01}) by
$k_x^0(t)=k_x^0-eE_xt.$ With this observation, the Hall conductance
reduces to,
\begin{equation}
\sigma_H=\frac{e^2}{h}\int\frac{dk_x^0dk_y}{2\pi}
\frac{\sin\theta\cos\theta}{1+\cos^2\theta}
\frac{\partial\theta}{\partial k_x^0}
\frac{\partial\varphi}{\partial k_y}.
\end{equation}
An interesting observation is that the correction of the decoherence
to the Hall conductance is zero, this can be understood by examining
Eq. (\ref{shenhallf}), keeping in mind that $\theta$ depends only on
$k_x^0$ while $\varphi$ only on $k_y$. It is important to point out
that the contribution from the steady state in the absence of
external field was ignored in this section, this is reasonable that
there has no current in the system when it reaches its steady state
without external driving fields. In other words, we here only have
interests in the current induced by the external fields, all of
other contributions do not concern us. The dependence of the Hall
conductivity on $\delta$ and $t_a$ is shown in Fig.~4.
 We find that $\sigma_H$ change sharply around
$t_a=0$ except at $\delta=0$, but  there is no phase transition at
$t_a=0$ in the sense that the Hall conductance has a same sign for
both positive and negative $t_a$. The topological phase changes with
the parity of $m$, when $m$ is an odd integer, $\sigma_H<0$, whereas
for even $m$, $\sigma_H>0$.

\section*{Discussion}

We have studied the Hall conductance of topological insulators in
the presence of decoherence. After extending the Bloch's theorem
from closed to open system, we have developed  an approach to
calculate perturbatively the steady state of the system  driven by a
perturbation. Then we apply this approach to derive the Hall
conductance for the open system. We expand the Hall conductance in
powers of dissipation rate, and find that the zeroth order
covers the usual Hall conductance when the open system decays from a
band to the others, whereas it can not return to the usual Hall
conductance with a dissipator in the other form. The first order
gives the correlation of the decoherence to the conductance, which
vanishes for the two-dimensional lattice and contributes non-zero
value to bulk ${\rm Bi_2Se_3}.$

Generally speaking, the Hall conductance for open system can not be
written as a multiple of a Chern number and a constant, or as a
weighted sum of Chern numbers, in this sense, there is no
topological invariant for open systems. The situation changes when a
dissipator keeps the density matrix of the steady state in a
diagonal form in a Hilbert space spanned by the instantaneous
eigenstates of the Hamiltonian. Specifically, when the steady state
takes, $\rho_{ss}(\vec k)=\sum_n
\alpha_{n,\vec{k}}|u_{n,\vec{k}}^{\prime}(t)\rangle\langle
u_{n,\vec{k}}^{\prime}(t)|$ with $\alpha_{n,\vec{k}}$ independent of
time, and  $e^{i\vec{k}\vec{r}}|u_{n,\vec{k}}^{\prime}(t)\rangle$
denotes a wavefunction subject to the Hamiltonian, the Hall
conductivity can be written as a weighted sum of Chern numbers. This
is easy to find by expanding $|u_{n,\vec{k}}^{\prime}(t)\rangle$  up
to first order in the field strength and substituting the expansion
into the Hall conductivity.

An interesting observation of this paper is that by properly
designing the Hamiltonian, the decoherence effect on the Hall
conductance can be eliminated in the two-band model. This
observation makes the TIs immune to influences  of  environment and
then support its application into quantum information processing.

The Kubo formula derived within the framework of linear response
theory applies for  equilibrium systems. Complementarily, we develop
a formalism to explore the linear response of an open system to
external field. Though we adopt a specific master equation to
develop the idea, the general conclusion in this paper should be
applicable to other open systems described by various master
equations, in particular, for a system not in its equilibrium state.
\\

This work is supported by the NSF of China under Grants No
11175032.

\section*{Methods}
\subsection*{Perturbation expansion of the steady state.}
We start with the master equation Eq.(\ref{mse}), and introduce a
perturbed term $\lambda H^{\prime}$ to the Hamiltonian,
\begin{equation}
H=H_0+\lambda H^{\prime}.
\end{equation}
When applying the perturbation theory, we may separate the total
Hamiltonian $H$  in such a way that $H_0$ is a proper Hamiltonian
easy for obtaining   the zeroth order steady state, while keep the
perturbation part $\lambda H^{\prime}$ small. The steady state
$\rho_{ss}$ can be given by solving
\begin{equation}
-\frac{i}{h}\left[ {H,\rho_{ss} } \right] + {\cal L} (\rho_{ss})=0.
\end{equation}
Up to  first order in $\lambda$, the steady state can be expressed
as,
\begin{equation}
\rho_{ss}= \rho_{ss}^{(0)}+ \lambda\rho_{ss}^{(1)}.
\end{equation}
The zeroth order steady state $\rho_{ss}^{(0)}$ is then given by,
\begin{equation}
\frac{i}{\hbar}\left[ {H_0,\rho_{ss}^{(0)}} \right] = {\cal L}
(\rho_{ss}^{(0)}), \label{steady0}
\end{equation}
while the first order satisfies,
\begin{equation}
\frac{i}{\hbar}\left[ {H^{\prime},\rho_{ss}^{(0)}}
\right]+\frac{i}{\hbar}\left[ {H_0,\rho_{ss}^{(1)}} \right] = {\cal
L} (\rho_{ss}^{(1)}).\label{steady1}
\end{equation}
In a  Hilbert space spanned by the eigenstates $\{|\phi_i\rangle\}$
of Hamiltonian $H_0$,
$H_0|\phi_i\rangle=\varepsilon_i|\phi_i\rangle,$ the steady state
can be written as,
\begin{eqnarray}
{\rho _{ss}} &=& \sum\limits_{ij} {{\alpha _{ij}}} |{\phi _i}\rangle
\langle {\phi _j}| \nonumber\\
&\simeq& \sum\limits_{ij} {(\alpha _{ij}^{(0)} + \lambda \alpha
_{ij}^{(1)})} |{\phi _i}\rangle \langle {\phi _j}|\nonumber\\
&=&\rho^{(0)}_{ss}+\lambda\rho^{(1)}_{ss}.
\end{eqnarray}
Substituting this expansion into Eq. (\ref{steady0}) and Eq.
(\ref{steady1}), we obtain an equation for  the  coefficients
$\alpha_{ij}^{(1)},$
\begin{eqnarray}
\langle {\phi _i}|{\cal L} (\rho _{ss}^{(1)})|{\phi _j}\rangle &=&
\frac{i}{\hbar }(\sum\limits_\beta  {H^{\prime}_{i\beta }}
\alpha _{\beta j}^{(0)} - \alpha _{i\beta }^{(0)}H^{\prime}_{\beta j})\nonumber\\
 &+& \frac{i}{\hbar }\alpha _{ij}^{(1)}({\varepsilon _i} - {\varepsilon
 _j}), \label{1thalpha}
\end{eqnarray}
where
$H^{\prime}_{\alpha\beta}=\langle\phi_\alpha|H^{\prime}|\phi_{\beta}\rangle,$
and $\alpha_{ij}^{(1)}=\alpha_{ji}^{(1)*}.$ Assume the zeroth order
steady state is easy to derive, the steady state up to first order
in $\lambda$ can be given by solving Eq. (\ref{1thalpha}).

In order to derive the Hall conductance as a response to an external
field, we consider the following idealized model: an non-interacting
electron gas in an periodic potential $V(\vec{r})$. In the presence
of a constant electric field $\vec{E}$ and when the field can be
represented by a time-dependent vector potential, the system
Hamiltonian takes\cite{bohm03},
\begin{equation}
H_0\left (k(t)\right )=\frac{1}{2m}\left
(-i\hbar\nabla+\hbar\vec{k}(t)\right )^2+V(\vec{r}),
\end{equation}
with $\vec{k}(t)=\vec{k}-e\vec{E}t.$ Taken the electric field  in
the x-direction, the y-component of the velocity operator in such a
case is given by $v_y=\frac{1}{\hbar}\frac{\partial
H_0(\vec{k})}{\partial k_y}$ \cite{bohm03}. The y-component of the
average velocity in the steady state is,
\begin{equation}
\bar{v}_y=\frac{1}{\hbar}\sum_{ij}\alpha_{ij}
\langle\phi_j|\frac{\partial H_0(\vec{k})}{\partial
k_y}|\phi_{i}\rangle.
\end{equation}
Up to first order in the perturbation $\lambda$, $\bar{v}_y$ takes
\begin{equation}
{{\bar v}_y} = \frac{1}{\hbar }\sum\limits_{i \ne j} {(\alpha
_{ij}^{(0)} + \lambda\alpha _{ij}^{(1)})\langle {\phi
_j}|\frac{{\partial H_0}}{{\partial
k_y}}|{\phi_i}\rangle}.\label{vy}
\end{equation}
The Hall current density is given by,
\begin{equation}
j_H=- e\int {\frac{{dk_xd{k_y}}}{{{{(2\pi )}^2}}}}  \cdot {{\bar
v}_y},
\end{equation}
the Hall conductivity $\sigma_H$ is defined as the ratio of this
current density and the electric field $E_x.$

To calculate perturbatively the Hall current, we work in the weak
field  limit, $E_x\sim 0$, this allows to use the adiabatic
approximation to specify the perturbation Hamiltonian $H^{\prime}$
induced by the adiabatic change of  Hamiltonian $H_0(\vec{k}(t))$
and calculate the perturbed steady state. We expand the density
matrix in the basis of the energy eigenstates
$|\phi_i[\vec{k}(t)]\rangle$ (the eigenstates of $H_0(\vec{k})$) as,
\begin{equation}
\rho[\vec{k}(t)]=\sum_{ij}\alpha_{ij}[\vec{k}(t)]
|\phi_i[\vec{k}(t)]\rangle\langle\phi_j[\vec{k}(t)]|,
\end{equation}
substituting this expansion into
\begin{equation}
\dot \rho  =  - i[H_0,\rho ] + {\cal L} (\rho ),
\end{equation}
we have,
\begin{eqnarray}
{{\dot \alpha }_{ij}} &=&  - i{\left[ {H_0,\rho } \right]_{ij}} +
{[{\cal L} (\rho )]_{ij}} - \sum\limits_m {\langle {\phi_i}|{{\dot
\phi}_m}\rangle }
{\alpha _{mj}}\nonumber\\
& -& \sum\limits_n {{\alpha _{in}}} \langle {{\dot
\phi}_n}|{\phi_j}\rangle,
\end{eqnarray}
where for the sake of simplicity we shorten the notations as
$\alpha_{ij}=\alpha_{ij}[\vec{k}(t)]$ and
$|\phi_i\rangle=|\phi_i[\vec{k}(t)]\rangle$. Notice that
\begin{equation}
\sum_{n}\alpha_{in}\langle \dot{
\phi}_n|\phi_j\rangle=-\sum_{n}\alpha_{in}\langle
\phi_n|\dot{\phi}_j\rangle,
\end{equation}
we obtain the Hamiltonian with a perturbation term $H^{\prime}$,
\begin{equation}
H_{mn}=H_0^{mn}-i\hbar\langle\phi_{m}|\frac{\partial}{\partial
t}|\phi_n\rangle=H^{mn}_{0}+H_{mn}^{\prime},\label{effH}
\end{equation}
where,
$$H_0^{mn}=\langle\phi_m|H_0|\phi_n\rangle=\varepsilon_n\delta_{mn}.$$
The Hamiltonian in Eq. (\ref{effH}) is the total Hamiltonian, which
includes a part of zeroth order in $E_x$ and a term of first order
in $E_x$. In the following, we shall take $E_x$ small such that
Hamiltonian  $H^{\prime}$ proportional to $E_x$ can be treated
perturbatively.

\subsection*{The zero-order steady state for two-band model.}
Solving the Schr\"odinger equation, $h_S|\Phi_E\rangle =
E|\Phi_E\rangle$ with Hamiltonian Eq. (\ref{H01}), we can obtain the
eigenenergies,
\begin{eqnarray}
{E_1} &=& \sqrt {{\varepsilon ^2} + {\Delta ^2}} ,\nonumber\\
{E_2} &=& -E_1, \label{H02}
\end{eqnarray}
and the corresponding eigenstates,
\begin{equation}
|\Phi_{E_j}\rangle = \left( \begin{array}{l}
{\phi _1}(E_j)\\
{\phi _2}(E_j)
\end{array} \right),\ \ j=1,2,
\label{phie}
\end{equation}
where,
\begin{eqnarray}
{\phi _1}(E_j) &=& \frac{{\Delta {e^{ - i\varphi}}}}
{{\sqrt {{\Delta ^2} + (E_j-\varepsilon)^2}}},\nonumber\\
{\phi _2}(E_j) &=& \frac{{E_j - \varepsilon }}{{\sqrt {{\Delta ^2} +
(E_j-\varepsilon)^2}}}.\label{phi1}
\end{eqnarray}
For the sake of simplicity, we transform the formalism into a
Hilbert space spanned by the eigenstates of $h_S$. Introducing $ U =
\left( {\begin{array}{*{20}{c}}
{{\phi _1}({E_1})}&{{\phi _1}({E_2})}\\
{{\phi _2}({E_1})}&{{\phi _2}({E_2})}
\end{array}} \right)
$, we find that ${h_S} = U{H_{dia}}{U^\dag }$ with $H_{dia}= \left(
{\begin{array}{*{20}{c}}
{{E_1}}&0\\
0&{{E_2}}
\end{array}} \right).
$ Define $F = {U^\dag }\sigma_- U$ and
$\cos\theta=\frac{\varepsilon}{\sqrt{\varepsilon^2+\Delta^2}}$, the
elements of matrix $F = \left( {\begin{array}{*{20}{c}}
{{F_{11}}}&{{F_{12}}}\\
{{F_{21}}}&{{F_{22}}}
\end{array}} \right)$ can be expressed as,
\begin{eqnarray}
{F_{11}} &=& \frac{1}{2}\sin\theta e^{ -i\varphi}, \,
{F_{12}} = -\sin^2\frac{\theta}{2}e^{-i\varphi},\nonumber\\
{F_{21}} &=& \cos^2\frac{\theta}{2}e^{-i\varphi}, \,
{F_{22}} = -
\frac{1}{2}\sin\theta e^{-i\varphi}. \label{Fij}
\end{eqnarray}
Collecting all these results, the master equation can be re-written
as,
\begin{equation}
{{\dot \rho }} =  - i[\sum_{k_x^0,k_y}H_{dia},{\rho}]
+\sum_{k_x^0,k_y} \gamma(2F{\rho}{F^\dag } - {F^\dag }F{\rho} -
{\rho}{F^\dag }F). \label{rho0}
\end{equation}
The steady state $\tau^{(0)}=\rho(t\rightarrow \infty) = \left(
{\begin{array}{*{20}{c}}
{{\tau^{(0)}_{11}}}&{{\tau^{(0)}_{12}}}\\
{{\tau^{(0)}_{21}}}&{{\tau^{(0)}_{22}}}
\end{array}} \right)$ with fixed $k_x^0$ and $k_y$ can be given by solving,
\begin{equation}
i[{H_{dia}},\tau^{(0)}]= \gamma(2F\tau^{(0)}{F^\dag } - {F^\dag
}F\tau^{(0)} - \tau^{(0)}{F^\dag }F), \label{rhos}
\end{equation}
this gives rise to,
\begin{eqnarray}
{\tau^{(0)}_{11}}&=&\frac{\sin^2\frac \theta 2(\gamma ^2 + 2E_1^2 -
2E_1 ^2\cos\theta )}{\gamma^2 + 3E_1^2
+E_1 ^2\cos(2\theta )},\nonumber\\
{\tau^{(0)}_{12}}&=&\frac{\gamma (\gamma  - 2iE_1 )\sin\theta }
{2\left[\gamma^2 + 3E_1^2 + E_1^2\cos(2\theta )\right]},\nonumber\\
{\tau^{(0)}_{21}}&=&\tau^{(0)*}_{12},\\
{\tau^{(0)}_{22}}&=&1-\tau^{(0)}_{11}. \nonumber \label{coefficient}
\end{eqnarray}
Here $\tau^{(0)}$ denotes the steady state without perturbations. In
weak dissipation limit $\gamma\rightarrow 0$, we find
$\tau^{(0)}_{12}=\tau^{(0)}_{21}\rightarrow 0$, and
$\tau^{(0)}_{11}$ approaches $\frac{\sin^2\frac\theta
2(1-\cos\theta)}{1+\cos^2\theta}$. Obviously,  in this limit,
$\tau^{(0)}_{11}=0$ when $\Delta=0$, leading to the thermal state
(ground state) at zero temperature. This observation suggests that
the steady state under study is  in general different from the Gibbs
states, as a consequence, the Hall conductance would be different
from that given by the Kubo formula.

\end{document}